\documentclass[conference,letterpaper]{IEEEtran}

\usepackage[utf8]{inputenc}
\usepackage[T1]{fontenc}
\usepackage{url}
\usepackage{ifthen}
\usepackage{cite}
\usepackage[cmex10]{amsmath} % Use the [cmex10] option to ensure complicance
\usepackage{pdfsync}

\usepackage{enumerate}
\usepackage{graphicx}
\usepackage{amssymb}
\usepackage{mathtools}
\usepackage{xspace}
\usepackage{subfigure}
\usepackage{colonequals}
\usepackage[mathscr]{euscript}
\usepackage{amsthm}
\usepackage{mathrsfs}

\usepackage{xspace}

\usepackage[lined,linesnumbered,ruled,commentsnumbered]{algorithm2e} % for algorithm

\usepackage{epsfig}
\usepackage{epstopdf}

\usepackage{tikz}
\usepackage{pgfplots}

\usetikzlibrary{arrows,shapes,backgrounds,plotmarks,positioning}

\pgfplotsset{compat=newest}                         % move axis labels close to the tick label automatically
\pgfplotsset{plot coordinates/math parser=false}
\newlength\figureheight
\newlength\figurewidth

\newtheorem{theorem}{Theorem}%[section]

\newcommand{\argmin}{\arg\!\min}

\newcommand{\op}{\text}
\newcommand{\SEN}{\mathcal{S}}
\newcommand{\PUB}{\mathcal{X}}
\newcommand{\REL}{\hat{\mathcal{X}}}
\newcommand{\Sen}{S}
\newcommand{\Pub}{X}
\newcommand{\Rel}{\hat{X}}
\newcommand{\sen}{s}
\newcommand{\pub}{x}
\newcommand{\rel}{\hat{x}}
\newcommand{\Pat}{\mathcal{P}}
\newcommand{\TW}{\hat{W}}
\newcommand{\TY}{\hat{Y}}

\newcommand{\E}{\mathbb{E}}

\newcommand{\Real}{\mathbb{R}}
\newcommand{\RealP}{\mathbb{R}_+}

\newcommand{\Set}[1]{\{#1\}}

\newcommand{\Simplex}{\triangle}

\begin{document}
\title{A Submodularity-based Agglomerative Clustering Algorithm for the Privacy Funnel}

%%% Several authors with up to three affiliations:
\author{%
  \IEEEauthorblockN{Ni Ding}
  \IEEEauthorblockA{Data61, CSIRO\\
                    5/13 Garden St, Eveleigh \\
                    Sydney, Australia 2015\\
                    Email: ni.ding@data61.csiro.au}
  \and
  \IEEEauthorblockN{Parastoo Sadeghi}
  \IEEEauthorblockA{Research School of Electrical,\\
                    Energy and Materials Engineering (RSEEME)\\
                    The Australian National University,
                    Canberra, Australia 2601\\
                    Email: parastoo.sadeghi@anu.edu.au}
}

\maketitle

%%%%%%
%% Abstract:
%% If your paper is eligible for the student paper award, please add
%% the comment "THIS PAPER IS ELIGIBLE FOR THE STUDENT PAPER
%% AWARD." as a first line in the abstract.
%% For the final version of the accepted paper, please do not forget
%% to remove this comment!
%%
\begin{abstract}
    For the privacy funnel (PF) problem, we propose an efficient iterative agglomerative clustering algorithm based on the minimization of the difference of submodular functions (IAC-MDSF). For a data curator that wants to share the data $\Pub$ correlated with the sensitive information $\Sen$, the PF problem is to generate the sanitized data $\Rel$ that maintains a specified utility/fidelity threshold on $I(\Pub; \Rel)$ while minimizing the privacy leakage $I(\Sen ; \Rel)$.
    Our IAC-MDSF algorithm starts with the original alphabet $\REL \coloneqq \PUB$ and iteratively merges the elements in the current alphabet $\REL$ that minimizes the Lagrangian function $I(\Sen ; \Rel) - \lambda I(\Pub ; \Rel)$. We prove that the best merge in each iteration of IAC-MDSF can be searched efficiently over all subsets of $\REL$ by the existing MDSF algorithms. We show that the IAC-MDSF algorithm also applies to the information bottleneck (IB), a dual problem to PF.
    By varying the value of the Lagrangian multiplier $\lambda$, we obtain the experimental results on a heart disease data set in terms of the Pareto frontier: $I(\Sen ; \Rel)$ vs. $-I(\Pub ; \Rel)$. We show that our IAC-MDSF algorithm outperforms the existing iterative pairwise merge approaches for both PF and IB and is computationally much less complex.
\end{abstract}

%% The paper must be self-contained. However, if you are referring to
%% a full version for checking certain proofs, please provide the
%% publically accessible location below.  If the paper is completely
%% self-contained, you can remove the following line from your
%% submission.

%\textit{A full version of this paper is accessible at:}
%\url{http://isit2019.fr/}

\section{Introduction}

As the multi-party data-sharing becomes inevitable in the \emph{big data} era, privacy and security issues of data sharing also create major challenges for data curators and owners. More precisely, when releasing the useful data $\Pub$, we also need to restrict the leakage of the sensitive/private information $\Sen$, e.g., an individual's medical records, due to the inherent correlation between $\Sen$ and $\Pub$. Existing privacy preserving techniques in computer science such as the celebrated differential privacy (DP) \cite{DPbook2014} add noise to the useful data $X$ to ensure the indifference of the noisy released dataset to the presence or absence of an individual's records. It is for sure that, if we distort $\Pub$ to protect against malicious inference on $\Sen$, we also lose the fidelity/utility. However, the DP framework does not allow  \emph{joint} optimization of privacy  and utility.

\begin{figure}[tbp]
   \centering
   \scalebox{0.7}{\begin{tikzpicture}

\draw [blue!20,fill = blue!20] (1.5,0) ellipse (0.25 and 0.33);
\draw [->,>=latex, blue!20,line width = 7](1.5,0) -- (3.5,0);

\node at (-1.5,0) {\Large $S$};
\draw [orange,line width =1.5](-1.2,0) -- (-0.3,0);
\node at (0,0) {\Large $\Pub$};
\node at (1.5,0.05) {\Large $\Rel$};
\draw [->,>=latex, purple!20,line width = 3](0.3,0) -- (1.2,0);

\draw [blue,dashed] (-1.8,0.4) rectangle (1.8,-0.4);
\node at (0,0.6) {\textcolor{blue}{Data curator}};

\node at (4.8,0) { public domain};

\end{tikzpicture} }
   % where an .eps filename suffix will be assumed under latex,
   % and a .pdf suffix will be assumed for pdflatex
   \caption{Statistical inference framework \cite{PvsInfer2012}: a data curator wants to publish $\Pub$ that is correlated with the private data $\Sen$. The privacy funnel problem \cite{PF2014} is to generate the sanitized data $\Rel$ via transition probability $p(\rel | \pub)$ so as to minimize the privacy leakage $I(\Sen ; \Rel)$ and maintain a utility threshold on $I(\Pub ; \Rel)$.}
   \label{fig:sys}
 \end{figure}
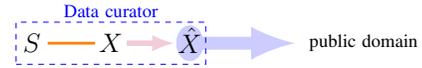

On the other hand, a general framework of statistical inference is outlined in \cite{PvsInfer2012}, which allows characterization of the \emph{privacy-utility tradeoff (PUT)}. The average privacy leakage is measured by the posterior knowledge gain $H(\Sen) - H(\Sen | \Rel) = I(\Sen ; \Rel)$ on $S$ at the adversary, where $\Rel$ is the released \emph{sanitized} data based on the original useful data $\Pub$. See Fig.~\ref{fig:sys}. The problem is to determine the transition probability $p(\rel | \pub)$ to minimize $I(\Sen ; \Rel)$ while ensuring the distortion in $\Rel$ remains below a certain level. It is shown in \cite[Theorem~1]{PvsInfer2012} that the solution $p(\rel | \pub )$ can be determined by solving a multivariate convex minimization problem, where the alternating minimization algorithms, e.g., the Blahut–Arimoto algorithm \cite{Blahut1972}, also apply.\footnote{The convexity holds if the distortion is measured by the expected distance between $\Pub$ and $\Rel$ \cite[Theorem~1]{PvsInfer2012}.}
But, this approach requires a given alphabet $\REL$ for the sanitization.

%%%%%%

In \cite{PF2014}, the authors used $I(\Pub ; \Rel)$ to measure the utility and considered the problem $\min_{p(\rel | \pub)} I(\Sen ; \Rel)$ s.t. $I(\Pub ; \Rel) \geq \theta_{\op{U}}$ for a guaranteed utility level $\theta_{\op{U}}$. It is called the \emph{privacy funnel (PF)} problem in that $p(\rel | \pub)$ behaves like a funnel to pass $\Pub$, but block $\Sen$. This problem is dual to the information bottleneck (IB) \cite{IB2000} in machine learning\footnote{IB tries to design the bottleneck $p(\rel | \pub)$ that passes relevant information about $\Sen$, which is nested in $\Pub$, with the minimum coding rate $I(\Pub ; \Rel)$ \cite{IB2000}. }
so that the idea of the agglomerative clustering \cite{IBAgglom1999} was borrowed in \cite[Algorithm~1]{PF2014} to produce a codebook $\REL$ from $\PUB$: through iteratively merging elements in $\Pub$ with a resulting $I(\Pub ; \Rel) \geq \theta_{\op{U}}$ that minimizes $I(\Sen ; \Rel)$.
Since the determination of the optimal merge incurs a combinatorial search, the authors in \cite{PF2014,IBAgglom1999} resorted to a brute-force pairwise merge approach so that the complexity is controlled at $O(|\PUB|^2)$ in each iteration.
However, it is well-known that some combinatorial optimization problems exhibiting strong structures can be solved efficiently in polynomial time \cite{Fujishige2005}.
Thus, it is worth understanding whether these techniques also apply to the PF problem.

In this paper, we propose a submodularity-based iterative agglomerative clustering approach, the IAC-MDSF algorithm, that starts with $\REL^{(0)} \coloneqq \PUB$ and iteratively searches an optimal merge over all subsets $W \subseteq \REL^{(k)}$ resulting in an r.v. $\Rel^{(k)}_W$ that minimizes the Lagrangian function $I (\Sen ; \Rel^{(k)}_W) - \lambda I (\Pub ; \Rel^{(k)}_W)$.
We prove that this minimization problem is reduced to the minimization of the difference of two submodular functions (MDSF), for which, the existing MDSF algorithms, e.g., \cite{SubMCover2013,SubMDiff2012,SubSuper2005}, can be applied to ensure a local convergence in polynomial time.
We show that this MDSF-based approach also applies to the IB problem.
We run experiments on the Hungarian heart disease data set in the UCI machine learning repository \cite{UCI2007}, where we vary the value of the Lagrangian multiplier $\lambda$ to outline the PUT as the Pareto frontier: the privacy leakage $I(\Sen ; \Rel)$ vs. utility loss $ - I(\Pub ; \Rel)$. We show that, for both PF and IB problems, our IAC-MDSF algorithm in general Pareto dominates or outperforms the pairwise merge approaches in \cite{PF2014} and is computationally much less complex.

%-------------------------------------------------------------------------system---------------------------------------------------------------------------

\section{Privacy against Statistical Inference Attack}

Consider the situation where a data curator wants to publish data $\Pub$ to the public. At the same time, he/she obtains some sensitive/private data $\Sen$ that is, in general, correlated with $\Pub$. In the public domain, there may exist some adversaries or legitimate, but curious users that can infer $S$ by observing $\Pub$. Thus, instead of the original $\Pub$, the data curator publishes a distorted version $\Rel$. The purpose is to keep the fidelity/utility of $\Pub$ in $\Rel$ while minimizing the leakage of $\Sen$ via $\Rel$.
We regard $\Sen$ and $\Pub$ as r.v.s with alphabets $\SEN$ and $\PUB$, respectively. The correlation between $\Sen$ and $\Pub$ is captured by the joint probability mass function $p(\sen,\pub)$. The design of the data release scheme $\Rel$ is to determine the mapping from $\PUB$ to $\REL$, or the transition probability $p(\rel | \pub)$ for all $(\pub,\rel) \in \PUB \times \REL$. Thus, there naturally arises a Markov chain $\Sen - \Pub - \Rel$.

\textbf{Privacy measure} \cite{PvsInfer2012}: Let $\Simplex_{\Sen}$ be the probability simplex over $\SEN$. For any user in the public domain, denote $q \in \Simplex_{\Sen}$ his/her belief on $S$ and $C(\Sen,q)$ the cost to infer $\Sen$ based on the distribution $q$.
It is assumed that any user in the public domain is able to adjust $q$ to minimize the expected prior inference cost $c^*_0 = \min_{q \in \Simplex_{\Sen}}  \E_{\Sen}[C(\Sen,q)]$ and posterior inference cost $c^*(\rel) = \min_{q \in \Simplex_{\Sen}} \E_{\Sen}[C(\Sen,q) | \Rel = \rel]$ (e.g., of a maximum a posteriori estimation). Thus, in order to preserve the privacy of $\Sen$, the goal is to remain the difficulty inferring $\Sen$, or minimize the average cost reduction $\delta C = c^*_0 - \E_{\Rel} [c^*(\rel)]$, in the public domain.
It is shown in \cite[Section~IV]{PvsInfer2012} that $\delta C = I(\Sen ; \Rel)$ if the log loss $C(\Sen,q) = - \log q(\sen)$ is adopted as the inference cost. Here, the mutual information $I(\Sen ; \Rel) = H(\Sen) - H(\Sen | \Rel)$ is interpreted as the leakage of $\Sen$ to the public via $\Rel$.

\subsection{Privacy Funnel}

\textbf{Utility measure}: The utility refers to how much useful information in $\Pub$ is revealed to the public via $\Rel$. It can be measured as the expected distortion $\E_{\Pub,\Rel}[d(\pub,\rel)]$ for some pairwise distance measure $d \colon \PUB \times \REL \mapsto \RealP$. Instead, the authors in \cite{PF2014} again considered the widely used log-loss in machine learning and information theory so that the utility is measured by $I(\Pub ; \Rel)$.
Thus, the optimization of PUT can be formulated as a constrained minimization problem, called the \emph{privacy funnel (PF)} \cite{PF2014}: for a given utility threshold $\theta_\op{U}$,
\begin{equation} \label{eq:PF}
    \begin{aligned}
        & \min_{p(\rel | \pub)} I (\Sen ; \Rel) \\
        & \op{s.t.} \quad  I (\Pub ; \Rel) \geq \theta_\op{U}.
    \end{aligned}
\end{equation}
This problem formulation also establishes the duality between PF and the information bottleneck (IB) problem \cite{IB2000} in machine learning \cite[Section~II]{PF2014}. See also Section~\ref{sec:IB}.
Although problem~\eqref{eq:PF} is not convex, %\footnote{The mutual  $I(\Sen ; \Rel)$ is convex in $p(\rel | \pub)$ \cite{PF2014} so that the constraint set in \eqref{eq:PF} is not convex. Problem~\eqref{eq:PF} is convex if replacing constraint $I (\Pub ; \Rel) \geq \theta_\op{U}$ by $\E_{\Pub,\Rel}[d(\pub,\rel)] \leq \theta_{\op{D}}$ for some distortion threshold \cite[Theorem~1]{PF2014}. This problem is solvable \cite[Remark~2]{PvsInfer2012} but requires a given $\REL$. }
it allows the agglomerative clustering algorithms \cite{IBAgglom1999} to not only search a deterministic solution $p(\rel | \pub)$, but also determine the alphabet $\REL$ from $\PUB$.

\section{Agglomerative Clustering Algorithm}

Consider the PF problem \eqref{eq:PF}. We have the Lagrangian function
\begin{equation} \label{eq:Lagrangian}
    L_{\op{PF}}(p(\rel | \pub),\lambda) =  I (\Sen ; \Rel) - \lambda I (\Pub ; \Rel).
\end{equation}
The solution of \eqref{eq:PF} for all $\theta_{\op{U}}$ can be determined if we solve $\min_{p(\rel | \pub)} L_{\op{PF}}(p(\rel | \pub),\lambda)$ for all $\lambda \geq 0$.\footnote{When $\lambda = 0$, \eqref{eq:PF} reduces to $\min_{p(\rel | \pub)} I (\Sen ; \Rel)$, where we only want to minimize the privacy leakage. }
Also, due to the PUT, $L_{\op{PF}}(p(\rel | \pub),\lambda)$ can be interpreted as a weighted sum of two conflicting objectives, for which, each $\lambda$ produces an achievable pair of mutual information $\Set{I(\Sen ; \Rel), -I(\Pub ; \Rel)}$ and all pairs form the \emph{Pareto frontier} indicating how best we can minimize the privacy and utility losses at the same time.
Thus, instead of \eqref{eq:PF}, it suffices to address how to solve the problem $\min_{p(\rel | \pub)} L_{\op{PF}}(p(\rel | \pub),\lambda)$ for any given $\lambda$.

%----------------------------------------------------------------clustering--------------------------------------------------------------------------------------------------

\subsection{Agglomerative Clustering Algorithm}

Let the alphabet $\REL$ be generated by a deterministic transition $p(\rel | \pub)$ or hard clustering of the elements in $\PUB$, i.e., the resulting $\REL = \Set{\TW \colon W \in \Pat}$ is built based on a partition $\Pat = \Set{W \colon  W \subseteq \PUB }$ of $\PUB$, where, for each $W \in \Pat$, all elements $\pub \in W$ are merged into the same element $\TW \in \REL$.
For example, for $\PUB = \Set{1,\dotsc,4}$, the partition $\Set{\Set{1,4},\Set{2,3}}$ yields $\REL = \Set{14,23}$, where $1$ and $4$ are merged to $14$ and $2$ and $3$ are merged to $23$.\footnote{In $\REL$, alphabet elements $\TW$, e.g., `$14$' and `$23$', denote the labels/indices of the merged elements, one can choose other labels based on the real application.}
The transition probability $p(\rel | \pub)$ is
$$ p(\TW|\pub) = \begin{cases} 1 & \pub \in W \\ 0 & \pub \notin W  \end{cases}, \qquad \forall W \in \Pat,$$
the resulting joint distribution $p(\sen,\rel)$ is
$$ p(\sen, \TW) = \sum_{\pub \in W} p(\sen,\pub), \qquad \forall \sen \in \Sen, W \in \Pat, $$
and the marginal distribution $p(\rel)$ is $p(\TW) = \sum_{\pub \in W} p(\pub), \forall W \in \Pat$.
For example, if $p(\Pub = 1) = 0.2$, $p(\Pub = 2) = 0.3$, $p(\Pub = 3) = 0.1$ and $p(\Pub = 4) = 0.4$, we have $p(\Rel = 14) = 0.6$ and $p(\Rel = 23) = 0.4$.

Instead of obtaining the partition $\Pat$ in a one-off manner, consider an iterative agglomerative clustering approach in Algorithm~\ref{algo:CombAggloCluster}: initiate $\REL^{(0)} \coloneqq \PUB$ and, in each iteration $k$, we obtain $W^*$ as the minimizer of
\begin{equation} \label{eq:LagrangianComb}
    \min \Set{ I (\Sen ; \Rel^{(k)}_W) - \lambda I (\Rel^{(k)} ; \Rel^{(k)}_W)  \colon W \subseteq \REL^{(k)} }
\end{equation}
and merge all $\rel^{(k)} \in W^*$ into $\TW^*$.
Let $\REL^{(k)}_W = (\REL^{(k)} \setminus W) \cup \Set{\TW}$ be the alphabet by merging all $\rel^{(k)} \in W$ into $\TW$. We have $\Rel^{(k)}_W$ in \eqref{eq:LagrangianComb} denote the resulting r.v..
For example, for $\REL^{(k)} = \Set{1,\dotsc,4}$ and $W = \Set{1,4}$, $\REL^{(k)}_W = \Set{14,2,3}$ and the resulting $\Rel^{(k)}_W$ has probabilities $p(\Rel^{(k)}_W = 14) = p(\Rel^{(k)} = 1) + p(\Rel^{(k)} = 4)$, $p(\Rel^{(k)}_W = 2) = p(\Rel^{(k)} = 2)$ and $p(\Rel^{(k)}_W = 3) = p(\Rel^{(k)} = 3)$. The iteration in Algorithm~\ref{algo:CombAggloCluster} terminates when there is no merge that reduces the objective function \eqref{eq:LagrangianComb}, i.e., when $W^* = \emptyset$ or $|W^*| = 1$.

Note, the basic idea of Algorithm~\ref{algo:CombAggloCluster} is proposed in \cite{IBAgglom1999,PF2014}.
The difference is that the algorithms in \cite{IBAgglom1999,PF2014} are iterative pairwise merge approaches in that $W^*$ is brute-force searched over $\Set{W \subseteq \REL^{(k)} \colon |W| = 2}$ each time, while Algorithm~\ref{algo:CombAggloCluster} searches $W^*$ over $2^{\REL^{(k)}}$ by solving \eqref{eq:LagrangianComb}, a minimization of a set function converted from the Lagrangian function \eqref{eq:Lagrangian}.
It is obvious that the constraint on pairwise combinations of $\REL^{(k)}$ in \cite{IBAgglom1999,PF2014} is to avoid dealing with set function optimization problem. However, in the next subsection, we show that problem \eqref{eq:LagrangianComb} can be converted to an MDSF problem, a local optimum of which can be searched efficiently.

%----------------------------------------------------------------MDSF--------------------------------------------------------------------------------------------------

\subsection{Minimizing Difference of Submodular Functions (MDSF)}

For a given alphabet, e.g., $\REL^{(k)}$ at any iteration $k$ of Algorithm~\ref{algo:CombAggloCluster}, define two set functions $f$ and $g$ as
\begin{equation}
    \begin{aligned}
        f(W) &= \sum_{\rel^{(k)} \in W} p(\rel^{(k)}) \log \frac{p(\rel^{(k)})}{p(\TW)},  \\
        g(W) &= \sum_{\sen \in \SEN} \sum_{\rel^{(k)} \in W} p(\sen,\rel^{(k)}) \log \frac{p(\sen,\rel^{(k)})}{p(s,\TW)},
    \end{aligned} \nonumber
\end{equation}
for all $W \subseteq \PUB^{(k)}$. The following result shows that we can decompose the objective function in \eqref{eq:LagrangianComb} into two submodular set functions. The proof of Theorem~\ref{theo:main} is in Appendix~\ref{app:theo:main}.

       \begin{algorithm} [t]
	       \label{algo:CombAggloCluster}
	       \small
	       \SetAlgoLined
	       \SetKwInOut{Input}{input}\SetKwInOut{Output}{output}
	       \SetKwFor{For}{for}{do}{endfor}
            \SetKwRepeat{Repeat}{repeat}{until}
            \SetKwIF{If}{ElseIf}{Else}{if}{then}{else if}{else}{endif}
	       \BlankLine
           \Input{$\lambda \in [0,1]$, $\SEN$, $\PUB$ and $p(\sen,\pub),\forall (\sen,\pub) \in \SEN \times \PUB$.}
	       \Output{alphabet $\REL$ and $p(\sen,\rel), \forall (\sen,\rel) \in \SEN \times \REL$.}
	       \BlankLine
                initiate $\REL^{(0)} \coloneqq \PUB$, $p(\sen,\rel^{(0)}) \coloneqq p(\sen,\pub)$ and $k \coloneqq 0$\;
                \Repeat{ $|W^*| \leq 1$ or $|\REL^{(k)}| = 1$}{
                    apply an MDSF algorithm to obtain the minimizer $W^*$ of \label{step:MDSF}
                        $$ \min \Set{ I (\Sen ; \Rel^{(k)}_W) - \lambda I (\Rel^{(k)} ; \Rel^{(k)}_W)  \colon W \subseteq \REL^{(k)} }; $$
                    \nl $\REL^{(k+1)} \coloneqq (\REL^{(k)} \setminus W^*) \cup \Set{\TW^*} $\;
                    \textbf{forall} $\sen \in \SEN$ \textbf{do} obtain
                    \begin{multline}
                        p(\sen, \rel^{(k+1)} ) \coloneqq \\ \begin{cases} p(\sen,\TW^*) & \rel^{(k+1)} = \TW^* \\
                                                                        p(\sen,\rel^{(k)}) & \rel^{(k+1)} \neq \TW^* \op{ and } \rel^{(k+1)} = \rel^{(k)}  \end{cases};  \nonumber
                    \end{multline}
                    \nl $k \coloneqq k + 1$\;
                }
                \Return $\REL^{(k)}$ and $p(\sen, \rel^{(k)})$\;
	   \caption{Iterative agglomerative clustering algorithm based on the minimization of the difference of submodular functions (IAC-MDSF)}
	   \end{algorithm}

\begin{theorem} \label{theo:main}
    In each iteration $k$ of Algorithm~\ref{algo:CombAggloCluster},
    \begin{multline} \label{eq:Eq}
        \argmin \Set{ I (\Sen ; \Rel^{(k)}_W) - \lambda I (\Rel^{(k)} ; \Rel^{(k)}_W)  \colon W \subseteq \REL^{(k)} } \\
            = \argmin \Set{ (1-\lambda) f (W) - g (W) \colon W \subseteq \REL^{(k)} },
    \end{multline}
    where $f$ and $g$ are submodular\footnote{A set function $f \colon 2^V \mapsto \Real$ is submodular if $f(X) + f(Y) \geq f(X \cup Y) + f(X \cap Y), \forall X, Y \subseteq V$; $-f$ is supermodular if $f$ is submodular \cite{Fujishige2005}. } and nonincreasing: $f(W) \geq f(Y)$ and $g(W) \geq g(Y)$ for all $Y \subseteq W$. \hfill \IEEEQED
\end{theorem}

Then, in order to determine $W^*$ in step~\ref{step:MDSF} of the IAC-MDSF algorithm in Algorithm~\ref{algo:CombAggloCluster}, we just need to solve the problem
\begin{equation} \label{eq:MDSF_PF}
    \min \Set{ (1-\lambda) f (W) - g (W) \colon W \subseteq \REL^{(k)} }.
\end{equation}
Since $f$ and $g$ are nonincreasing, for all $\lambda \geq 1$, we have the minimizer of \eqref{eq:MDSF_PF} being the empty set $\emptyset$, i.e., Algorithm~\ref{algo:CombAggloCluster} just returns $\REL = \PUB$ and $p(\sen,\rel) = p(\sen,\pub), \forall \sen, \rel = \pub$. Then, to determine the Pareto frontier, we only need to solve the problem~\eqref{eq:MDSF_PF} for all $\lambda \in [0,1]$. In this case, $(1-\lambda) f$ and $g$ are both submodular and the problem \eqref{eq:MDSF_PF} is an MDSF.
The MDSF problem arises in many machine learning applications, e.g., feature selection, discriminative structured graph learning \cite{SubMCover2013}, for which, there are many polynomial time algorithms proposed in the literature, e.g., the \cite{SubMCover2013,SubMDiff2012}, that ensure convergence to a local optimum.

 \begin{figure*}[t]%[htbp]
   \centerline{
        \subfigure[$\Sen = \Set{\op{`age', } \op{`sex'}}$, $\Pub = \Set{\op{`sex', } \op{`cholesterol'} }$ and $\lambda = 0.8$.]{\scalebox{0.6}{% This file was created by matlab2tikz v0.4.3.
% Copyright (c) 2008--2013, Nico Schlömer <nico.schloemer@gmail.com>
% All rights reserved.
%
% The latest updates can be retrieved from
%   http://www.mathworks.com/matlabcentral/fileexchange/22022-matlab2tikz
% where you can also make suggestions and rate matlab2tikz.
%
\begin{tikzpicture}

\begin{axis}[%
width=4.2in,
height=1.3in,
scale only axis,
xmin=1,
xmax=12,
xlabel={\large iteration index $k$},
ymin=-1.8,
ymax=0.2,
ylabel={$I(\Sen;\Rel^{(k)}) - \lambda I(\Pub;\Rel^{(k)})$},
legend style={draw=darkgray!60!black,fill=white,legend cell align=left}
]
\addplot [
color=red,
solid,
line width = 1.5pt,
mark=o,
mark options={solid},
]
table[row sep=crcr]{
1 -1.05665196344758\\
2 -1.27106641606008\\
3 -1.4034258888698\\
4 -1.51696653151415\\
5 -1.59890655217136\\
6 -1.65183516490272\\
7 -1.67973962069092\\
8 -1.6817627961425\\
9 -1.69095311927469\\
10 -1.6934086570124\\
11 -1.71141164575166\\
12 -1.72604613544568\\
};
\addlegendentry{\large IAC-MDSF algorithm for PF};

\addplot [
color=blue,
solid,
line width = 1.5pt,
mark=asterisk,
mark options={solid},
]
table[row sep=crcr]{
1 -1.05665196344758\\
2 0.125586078349769\\
3 0.163554471226043\\
4 0.167122574966068\\
};
\addlegendentry{\large IAC-MDSF algorithm for IB};

\end{axis}
\end{tikzpicture}% 
}}
     \qquad\qquad
     \subfigure[$\Sen = \Set{\op{`age', } \op{`sex'}}$, $\Pub = \Set{\op{`age', } \op{`cholesterol'} }$ and $\lambda = 0.7$.]{\scalebox{0.6}{% This file was created by matlab2tikz v0.4.3.
% Copyright (c) 2008--2013, Nico Schlömer <nico.schloemer@gmail.com>
% All rights reserved.
%
% The latest updates can be retrieved from
%   http://www.mathworks.com/matlabcentral/fileexchange/22022-matlab2tikz
% where you can also make suggestions and rate matlab2tikz.
%
\begin{tikzpicture}

\begin{axis}[%
width=4.2in,
height=1.3in,
scale only axis,
xmin=1,
xmax=11,
xlabel={\large iteration index $k$},
ymin=-1.3,
ymax=0.8,
ylabel={$I(\Sen;\Rel^{(k)}) - \lambda I(\Pub;\Rel^{(k)})$},
legend style={at={(0.65,0.98)},anchor=north west,draw=black,fill=white,legend cell align=left}
]

\addplot [
color=red,
solid,
line width = 1.5pt,
mark=o,
mark options={solid},
]
table[row sep=crcr]{
1 -0.0907134342516134\\
2 -0.592210123547721\\
3 -0.834545008769874\\
4 -0.94611798554763\\
5 -1.04418685861908\\
6 -1.07808408407711\\
7 -1.13247486867793\\
8 -1.16158642481363\\
9 -1.18607622073199\\
10 -1.20190252175682\\
11 -1.20394333808335\\
};
\addlegendentry{\large IAC-MDSF algorithm for PF};

\addplot [
color=blue,
solid,
line width = 1.5pt,
mark=asterisk,
mark options={solid},
]
table[row sep=crcr]{
1 -0.0907134342516134\\
2 0.469646804170453\\
3 0.584030530292665\\
4 0.615864813716132\\
5 0.648428205906471\\
6 0.65974936662591\\
7 0.676269239085196\\
};
\addlegendentry{\large IAC-MDSF algorithm for IB};

\end{axis}
\end{tikzpicture}% 
}}
       }
   \caption{ The convergence of the Lagrangian function $I(\Sen;\Rel^{(k)}) - \lambda I(\Pub;\Rel^{(k)})$ when the IAC-MDSF algorithm in Algorithm~\ref{algo:CombAggloCluster} is applied to the PF and IB problems on the Hungarian heart disease data set in \cite{UCI2007}. }
   \label{fig:ObjConverge_Health}
 \end{figure*}
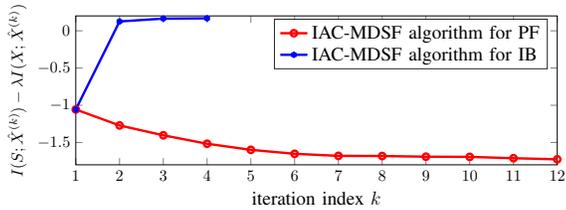
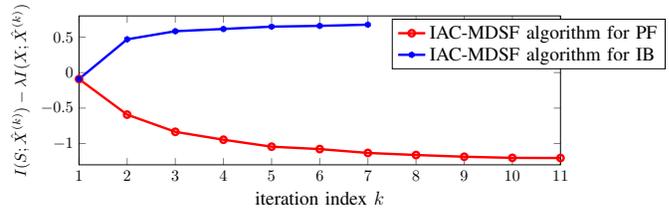

  \begin{figure*}[htbp]
   \centerline{\subfigure[$\Sen = \Set{\op{`age', } \op{`sex'}}$ and $\Pub = \Set{\op{`sex', } \op{`cholesterol'} }$.]{\scalebox{0.6}{% This file was created by matlab2tikz v0.4.3.
% Copyright (c) 2008--2013, Nico Schlömer <nico.schloemer@gmail.com>
% All rights reserved.
%
% The latest updates can be retrieved from
%   http://www.mathworks.com/matlabcentral/fileexchange/22022-matlab2tikz
% where you can also make suggestions and rate matlab2tikz.
%
\begin{tikzpicture}

\begin{axis}[%
width=4in,
height=1.7in,
scale only axis,
xmin=-1,
xmax=0.05,
xlabel={utility loss $-I(\Pub;\Rel)/H(\Pub)$},
ymin=0,
ymax=0.9,
ylabel={privacy leakage $I(\Sen;\Rel)/H(\Sen)$},
legend style={at={(0.65,0.98)},anchor=north west,draw=black,fill=white,legend cell align=left}
]

\addplot [
color=red,
only marks,
mark=asterisk,
mark options={solid},
]
table[row sep=crcr]{
1.74732878717024e-16 2.27647895327788e-16\\
1.74732878717024e-16 2.27647895327788e-16\\
1.74732878717024e-16 2.27647895327788e-16\\
1.74732878717024e-16 2.27647895327788e-16\\
1.74732878717024e-16 2.27647895327788e-16\\
1.74732878717024e-16 2.27647895327788e-16\\
1.74732878717024e-16 2.27647895327788e-16\\
1.74732878717024e-16 2.27647895327788e-16\\
1.74732878717024e-16 2.27647895327788e-16\\
1.74732878717024e-16 2.27647895327788e-16\\
1.74732878717024e-16 2.27647895327788e-16\\
1.74732878717024e-16 2.27647895327788e-16\\
1.74732878717024e-16 2.27647895327788e-16\\
1.74732878717024e-16 2.27647895327788e-16\\
1.74732878717024e-16 2.27647895327788e-16\\
1.74732878717024e-16 2.27647895327788e-16\\
1.74732878717024e-16 2.27647895327788e-16\\
1.74732878717024e-16 2.27647895327788e-16\\
1.74732878717024e-16 2.27647895327788e-16\\
1.74732878717024e-16 2.27647895327788e-16\\
1.74732878717024e-16 2.27647895327788e-16\\
1.74732878717024e-16 2.27647895327788e-16\\
1.74732878717024e-16 2.27647895327788e-16\\
1.74732878717024e-16 2.27647895327788e-16\\
1.74732878717024e-16 2.27647895327788e-16\\
1.74732878717024e-16 2.27647895327788e-16\\
1.74732878717024e-16 2.27647895327788e-16\\
1.74732878717024e-16 2.27647895327788e-16\\
1.74732878717024e-16 2.27647895327788e-16\\
1.74732878717024e-16 2.27647895327788e-16\\
1.74732878717024e-16 2.27647895327788e-16\\
1.74732878717024e-16 2.27647895327788e-16\\
1.74732878717024e-16 2.27647895327788e-16\\
1.74732878717024e-16 2.27647895327788e-16\\
1.74732878717024e-16 2.27647895327788e-16\\
1.74732878717024e-16 2.27647895327788e-16\\
1.74732878717024e-16 2.27647895327788e-16\\
1.74732878717024e-16 2.27647895327788e-16\\
1.74732878717024e-16 2.27647895327788e-16\\
1.74732878717024e-16 2.27647895327788e-16\\
-0.0415762151943791 0.0188096565117947\\
1.74732878717024e-16 -2.27647895327788e-16\\
1.74732878717024e-16 -2.27647895327788e-16\\
1.74732878717024e-16 -2.27647895327788e-16\\
-0.0214080551471489 0.0106608711652433\\
-0.0504412084501375 0.0240157049570442\\
-0.0358518886764433 0.015236294415251\\
-0.0468713248611433 0.0192906831015894\\
-0.0820876871568776 0.0330087134088065\\
-0.0970757004621353 0.0369698473812844\\
-0.0814850867047072 0.03362509418761\\
-0.108222159699337 0.0445445688678818\\
-0.120524837347667 0.0512099435874396\\
-0.118600305844046 0.0473071543808331\\
-0.164555172844703 0.0622800695100028\\
-0.18174871642703 0.0702507408333278\\
-0.239914589743708 0.111033367062478\\
-0.18257197388632 0.0718845601046519\\
-0.225231690867144 0.0920301852992658\\
-0.219734695559466 0.0835839881203125\\
-0.23366849564207 0.0768313264557618\\
-0.223707766706294 0.0786924562886372\\
-0.255796710401339 0.0945633098131916\\
-0.327815808266077 0.145242828919708\\
-0.281811223208344 0.094579236322601\\
-0.295555897168505 0.100400568986486\\
-0.315163791932383 0.12049602096133\\
-0.31003736346888 0.106049207096596\\
-0.36154515053787 0.130920788287992\\
-0.351941137105822 0.120589605534613\\
-0.369858117483862 0.139493995301604\\
-0.359723833175251 0.123181513723844\\
-0.380568604024083 0.143483189242069\\
-0.400088838675473 0.145632130977783\\
-0.389441322959048 0.137506798862212\\
-0.430000407479031 0.139884869806557\\
-0.471074656530631 0.194150190950885\\
-0.486806592497187 0.203312998022985\\
-0.451881682994052 0.157059244112841\\
-0.491267554874935 0.209124890460673\\
-0.48793038356961 0.207194249532939\\
-0.4826133221 0.198306030658267\\
-0.476866073089555 0.185527876537567\\
-0.511209600936371 0.226189909693816\\
-0.591074017046942 0.326614260294551\\
-0.551316916673667 0.271191659460535\\
-0.559034450778571 0.281246323522627\\
-0.511285274650854 0.219037084317485\\
-0.560682709123011 0.282185160882864\\
-0.558771367296646 0.279695000203211\\
-0.575290510800813 0.301216698231385\\
-0.538680284615922 0.253519659187223\\
-0.584935677117137 0.313782746636926\\
-0.591189253500679 0.321930117140048\\
-0.569577905276614 0.293774122833038\\
-0.571855144267731 0.296740986695708\\
-0.601584578162838 0.335473497172275\\
-0.605897229012965 0.341092164343123\\
-0.553396505768328 0.272692449088708\\
-0.560235561163807 0.281602601420902\\
-1 0.85454256370009\\
};
\addlegendentry{IAC-MDSF algorithm for PF};

\addplot [
color=green,
only marks,
line width = 1.5pt,
mark=diamond,
mark options={solid},
]
table[row sep=crcr]{
-0.0880535472753068 0.0548715052610257\\
-0.239166586738342 0.161299772481961\\
-0.358249930025776 0.256983864317453\\
-0.507000043972533 0.393041818950645\\
-0.546591214074366 0.424482350549958\\
-1 0.85454256370009\\
};
\addlegendentry{pairwise merge \cite[Algorithm~1]{PF2014} for PF};

\addplot [
color=blue,
only marks,
mark=o,
mark options={solid},
]
table[row sep=crcr]{
-1 0.85454256370009\\
-1 0.85454256370009\\
-1 0.85454256370009\\
-1 0.85454256370009\\
-1 0.85454256370009\\
-1 0.85454256370009\\
-1 0.85454256370009\\
-0.999072353433522 0.85454256370009\\
-1 0.85454256370009\\
-1 0.85454256370009\\
-1 0.85454256370009\\
-1 0.85454256370009\\
-1 0.85454256370009\\
-1 0.85454256370009\\
-1 0.85454256370009\\
-1 0.85454256370009\\
-1 0.85454256370009\\
-1 0.85454256370009\\
-1 0.85454256370009\\
-1 0.85454256370009\\
-1 0.85454256370009\\
-1 0.85454256370009\\
-1 0.85454256370009\\
-1 0.85454256370009\\
-1 0.85454256370009\\
-1 0.85454256370009\\
-1 0.85454256370009\\
-1 0.85454256370009\\
-0.999072353433522 0.85454256370009\\
-1 0.85454256370009\\
-1 0.85454256370009\\
-1 0.85454256370009\\
-1 0.85454256370009\\
-1 0.85454256370009\\
-1 0.85454256370009\\
-1 0.85454256370009\\
-1 0.85454256370009\\
-1 0.85454256370009\\
-1 0.85454256370009\\
-1 0.85454256370009\\
-1 0.85454256370009\\
-1 0.85454256370009\\
-1 0.85454256370009\\
-1 0.85454256370009\\
-1 0.85454256370009\\
-1 0.85454256370009\\
-1 0.85454256370009\\
-1 0.85454256370009\\
-1 0.85454256370009\\
-1 0.85454256370009\\
-1 0.85454256370009\\
-1 0.85454256370009\\
-0.998144706867044 0.853333994425565\\
-1 0.85454256370009\\
-0.952836744290039 0.83618397010523\\
-1 0.85454256370009\\
-0.999072353433522 0.85454256370009\\
-0.691800342811888 0.658959666520141\\
-0.672287818119213 0.643730252219361\\
-0.556543071639952 0.56758598187031\\
-0.525388122866844 0.541635397710626\\
-0.592913767424977 0.579100117824331\\
-0.460975676691742 0.495227181925909\\
-0.387867041786425 0.434416294001218\\
-0.334434811914907 0.389243657570685\\
-0.385536081378814 0.425805887451428\\
-0.370079865096918 0.41397707827126\\
-0.366711648091028 0.41509155615038\\
-0.303344705731958 0.360667714132805\\
-0.287719194620189 0.346056881503406\\
-0.293157414871549 0.347219826468119\\
-0.26611970051645 0.325563566542874\\
-0.276956227642858 0.324030706646002\\
-0.213166061795608 0.249121435539428\\
-0.17441256249986 0.215019914050942\\
-0.174028281781698 0.210402423669396\\
-0.141392155931488 0.177711462434244\\
-0.138836593999073 0.175590559318881\\
-0.148890193679961 0.184090339671257\\
-0.126252470523079 0.1604041075482\\
-0.131381729886311 0.165421943672808\\
-0.127180117089557 0.161612676822725\\
-0.123630953974921 0.158197276532081\\
-0.121901849002216 0.155944540179727\\
-0.0901863925041286 0.116289308580619\\
-0.0889086115379212 0.1158331416602\\
-0.0845955836218743 0.110213983235138\\
-0.075439946818584 0.0982857104111667\\
-0.075439946818584 0.0982857104111667\\
-0.075439946818584 0.0982857104111667\\
-0.0658971472231419 0.0846444595521101\\
-0.0646193662569345 0.0841882926316912\\
-0.0646193662569345 0.0841882926316912\\
-0.0535021029229107 0.0697043465170398\\
-0.0535021029229107 0.0697043465170398\\
-0.0535021029229107 0.0697043465170398\\
-0.0535021029229107 0.0697043465170398\\
-0.0535021029229107 0.0697043465170398\\
-0.0535021029229107 0.0697043465170398\\
-0.0535021029229107 0.0697043465170398\\
-0.0535021029229107 0.0697043465170398\\
};
\addlegendentry{IAC-MDSF algorithm for IB};

\addplot [
color=black,
only marks,
mark=square,
line width = 1.5pt,
mark options={solid},
]
table[row sep=crcr]{
-0.967402186602242 0.85454256370009\\
-0.533540233418074 0.538201651223979\\
-0.317572250584384 0.332774818242364\\
-0.156862103471918 0.167566367821508\\
-0.0795663956266259 0.0843499645551739\\
-0.0672001846560315 0.0701998189097965\\
};
\addlegendentry{pairwise merge \cite[Algorithm~2]{PF2014} for IB};

\end{axis}
\end{tikzpicture}% 
}}
     \qquad
     \subfigure[$\Sen = \Set{\op{`age', } \op{`sex'}}$ and $\Pub = \Set{\op{`age', } \op{`cholesterol'} }$.]{\scalebox{0.6}{\input{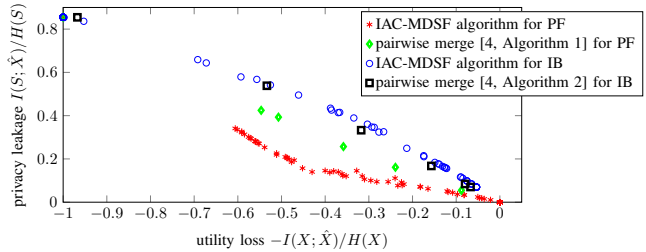}
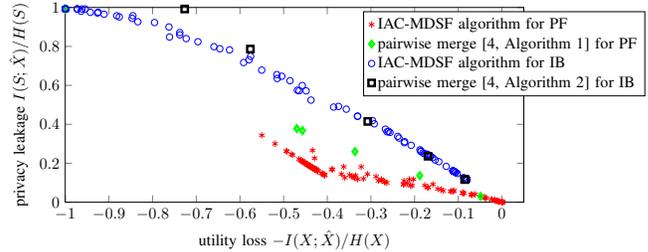}}}
   \caption{The Pareto frontiers of the PF and IB problems obtained by applying the IAC-MDSF algorithm in Algorithm~\ref{algo:CombAggloCluster} for multiple values of $\lambda \in [0,1]$ to the Hungarian heart disease data set in \cite{UCI2007}. Note, for IB, the Pareto frontier is interpreted as the extracted useful information on $\Sen$ vs. the reduction in coding rate. The results are compared with the iterative pairwise merge algorithms \cite[Algorithms~1]{PF2014} for PF and \cite[Algorithms~2]{PF2014} for IB. }
   \label{fig:PvsU_Health}
 \end{figure*}

%----------------------------------------------------------------IB--------------------------------------------------------------------------------------------------

\subsection{Information Bottleneck Problem}
\label{sec:IB}

The duality relationship between PF and information bottleneck (IB) has been pointed out in \cite{PF2014,GIB2018}. In IB \cite{IB2000}, $\Sen$ refers to the useful/relevent signal
that is hidden in the observations $\Pub$. The problem is to encode $\Pub$ into $\Rel$ with the minimum rate $I(\Pub ; \Rel)$ that extracts out the most information on $\Sen$, e.g., modeling speech phonemes from the audio waves.
The optimization is exactly the opposite of PF: given a coding rate threshold $\theta_\op{R}$,
\begin{equation} \label{eq:IB}
    \begin{aligned}
        & \max_{p(\rel | \pub)} I (\Sen ; \Rel) \\
        & \op{s.t.} \quad  I (\Pub ; \Rel) \leq \theta_\op{R}.
    \end{aligned}
\end{equation}
The Lagrangian function is\footnote{The original IB problem in \cite{IB2000} is formulated as minimizing the coding rate $I(\Pub ; \Rel)$ subject to the relevance $I(\Sen ; \Rel)$ is no less than some threshold. This problem and \eqref{eq:IB} share the same Lagrangian function $L_{\op{IB}}(p(\rel | \pub),\lambda)$. }
    $$ L_{\op{IB}}(p(\rel | \pub),\lambda) =  -I (\Sen ; \Rel) + \lambda I (\Pub ; \Rel) = - L_{\op{PF}}(p(\rel | \pub),\lambda)$$
and the Pareto frontier for IB can be outlined by maximizing $L_{\op{PF}}(p(\rel | \pub),\lambda)$ for all $\lambda \geq 0$. It is also obvious that, if we determine $W^*$ as the minimizer of
\begin{equation} \label{eq:MDSF_IB}
    \min \Set{ g (W) - (1-\lambda) f (W) \colon W \subseteq \REL^{(k)} }
\end{equation}
in step~\ref{step:MDSF},\footnote{This is equivalent to replacing the minimization problem in step~\ref{step:MDSF} of Algorithm~\ref{algo:CombAggloCluster} by $ \max \Set{ I (\Sen ; \Rel^{(k)}_W) - \lambda I (\Rel^{(k)} ; \Rel^{(k)}_W)  \colon W \subseteq \REL^{(k)} } $. }
Algorithm~\ref{algo:CombAggloCluster} returns a hard clustering solution and corresponding codebook $\REL$ for the IB problem.
For \eqref{eq:MDSF_IB}, we just need to consider $\lambda \in [0, 1]$ since the minimizer is $\REL^{(k)}$ for all $\lambda \geq 1$. Again, \eqref{eq:MDSF_IB} for all $\lambda \in [0,1]$ is an MDSF problem. This means that, for the same $\lambda$, the IAC-MDSF algorithm in Algorithm~\ref{algo:CombAggloCluster} can provide solutions for both PF and IB problems.

%----------------------------------------------------------------experiment--------------------------------------------------------------------------------------------------

\section{Experimental Results}

The UCI machine learning repository \cite{UCI2007} contains 463 data sets. In this repository, we use the heart disease data set created by the Hungarian Institute of Cardiology that contains patients data including 76 attributes to identify the presence of heart disease. We extract three of them, `age', `sex' and `serum cholesterol (mg/dl)' to run the following experiments based on two settings: the first is when $\Sen = \Set{\op{`age', } \op{'sex'}}$ and $\Pub = \Set{\op{'sex', } \op{`cholesterol'} }$; the second is when $\Sen = \Set{\op{`age', } \op{`sex'}}$ and $\Pub = \Set{\op{`age', } \op{`cholesterol'} }$.
For solving problems~\eqref{eq:MDSF_PF} and \eqref{eq:MDSF_IB} in step~\ref{step:MDSF} of Algorithm~\ref{algo:CombAggloCluster} for PF and IB, respectively, we run the function \verb"sfo_ssp" in the SFO toolbox \cite{SFOtools2010}. This function implements the submodular-supermodular (SubM-SuperM) algorithm proposed in \cite[Algorithm~1]{SubSuper2005} for solving MDSF problems.

\subsection{Convergence}

We first show the convergence performance of the IAC-MDSF algorithm for both PF and IB problems in Fig.~\ref{fig:ObjConverge_Health}, which is consistent with \cite[Lemma~3.3]{SubMDiff2012} that the SubM-SuperM algorithm ensures a reduction for PF and an increase for IB of the Lagrangian function in each iteration.

%\begin{figure}[tbp]
%	\centering
%    \scalebox{0.6}{\input{figures/ObjConverge_HealthDoubleSex.tex}}
%	\caption{The convergence of the Lagrangian function $I(\Sen;\Rel^{(k)}) - \lambda I(\Pub;\Rel^{(k)})$ when Algorithm~\ref{algo:CombAggloCluster} is applied to PF and IB problem for $\lambda = 0.8$ on the Hungarian heart disease data set in \cite{UCI2007}. Here, $\Sen = \Set{\op{`age', } \op{`sex'}}$ and $\Pub = \Set{\op{`sex', } \op{`cholesterol'} }$. }
%	\label{fig:ObjConverge_HealthDoubleSex}
%\end{figure}
%
%\begin{figure}[tbp]
%	\centering
%    \scalebox{0.6}{\input{figures/ObjConverge_HealthDoubleAge.tex}}
%	\caption{The convergence of the Lagrangian function $I(\Sen;\Rel^{(k)}) - \lambda I(\Pub;\Rel^{(k)})$ when Algorithm~\ref{algo:CombAggloCluster} is applied to PF and IB problem for $\lambda = 0.7$ on the Hungarian heart disease data set in \cite{UCI2007}. Here, $\Sen = \Set{\op{`age', } \op{`sex'}}$ and $\Pub = \Set{\op{`age', } \op{`cholesterol'} }$. }
%	\label{fig:ObjConverge_HealthDoubleAge}
%\end{figure}

\subsection{Pareto Frontier}

In Fig.~\ref{fig:PvsU_Health}, we apply the IAC-MDSF algorithm to get the Pareto frontiers for PF and IB problems by varying $\lambda$ from $0$ to $1$.
The Pareto frontier is presented in terms of two normalized mutual information: $I(S;Y)/H(S)$ and $-I(X;Y)/H(X)$. For PF, this is respectively interpreted as the loss in privacy, the leakage of $\Sen$, vs. the loss in utility; For IB, this is respectively interpreted as the extracted useful information vs. the reduction in coding rate.

We also plot the Pareto frontiers obtained by the pairwise merge algorithms proposed in \cite{PF2014}. For PF, \cite[Algorithm~1]{PF2014} iteratively searches two elements in $W^* = \Set{i,j} \subseteq \REL^{(k)}$ with $I(\Pub ; \Rel^{(k)}_{W^*}) \geq \theta_{\op{U}}$ that minimizes $I(\Sen ; \Rel^{(k)}_{W^*})$ and merge them to form the new alphabet $\REL^{(k+1)}$; For IB, \cite[Algorithm~2]{PF2014} iteratively merges $W^* = \Set{i,j} \subseteq \REL^{(k)}$ with $I(\Sen ; \Rel^{(k)}_{W^*}) \geq \theta_{\op{U}}$ that minimizes $I(\Pub ; \Rel^{(k)}_{W^*})$.
It can be seen that the IAC-MDSF algorithm in general outperforms \cite[Algorithms~1 and 2]{PF2014}.

%
%\begin{figure}[tbp]
%	\centering
%    \scalebox{0.6}{\input{figures/PvsU_HealthDoubleSex.tex}}
%	\caption{Pareto frontier of PF and IB obtained by applying Algorithm~\ref{algo:CombAggloCluster} for multiple values of $\lambda \in [0,1]$ to the Hungarian heart disease data set in \cite{UCI2007}. Here, $\Sen = \Set{\op{`age', } \op{`sex'}}$ and $\Pub = \Set{\op{`sex', } \op{`cholesterol'} }$. Note, for IB, the Pareto frontier is interpreted as the extracted useful information on $\Sen$ vs. the reduction in coding rate.}
%	\label{fig:PvsU_HealthDoubleSex}
%\end{figure}
%
%\begin{figure}[tbp]
%	\centering
%    \scalebox{0.6}{\input{figures/PvsU_HealthDoubleAge.tex}}
%	\caption{Pareto frontier of PF and IB obtained by applying Algorithm~\ref{algo:CombAggloCluster} for multiple values of $\lambda \in [0,1]$ to the Hungarian heart disease data set in \cite{UCI2007}. Here, $\Sen = \Set{\op{`age', } \op{`sex'}}$ and $\Pub = \Set{\op{`age', } \op{`cholesterol'} }$. Note, for IB, the Pareto frontier is interpreted as the extracted useful information on $\Sen$ vs. the reduction in coding rate.}
%	\label{fig:PvsU_HealthDoubleAge}
%\end{figure}

\subsection{Complexity}
\label{sec:Complexity}

\begin{figure}[tbp]
	\centering
    \scalebox{0.6}{% This file was created by matlab2tikz v0.4.3.
% Copyright (c) 2008--2013, Nico SchlÃ¶mer <nico.schloemer@gmail.com>
% All rights reserved.
%
% The latest updates can be retrieved from
%   http://www.mathworks.com/matlabcentral/fileexchange/22022-matlab2tikz
% where you can also make suggestions and rate matlab2tikz.
%
\begin{tikzpicture}

\begin{axis}[%
width=4.5in,
height=1.3in,
scale only axis,
xmin=1,
xmax=153,
xlabel={\large iteration index $k$},
ymin=1,
ymax=8,
legend style={at={(0.4,0.98)},anchor=north west,draw=black,fill=white,legend cell align=left}
]

\addplot [
color=red,
solid,
line width = 1.5pt
]
table[row sep=crcr]{
1 4.80999711111437\\
2 4.74064187119972\\
3 4.69186623320957\\
4 4.64697465816338\\
5 4.60398566674041\\
6 4.55969603944927\\
7 4.51411752701136\\
8 4.47342371020467\\
9 4.43265026453525\\
10 4.39124580848457\\
11 4.34945638109918\\
12 4.30729651663657\\
13 4.26477969159054\\
14 4.2225044197375\\
15 4.17949277198356\\
16 4.1400149175788\\
17 4.09986767901461\\
18 4.06074761068701\\
19 4.02243145778668\\
20 3.983850022531\\
21 3.94501028766861\\
22 3.9075891493832\\
23 3.87069957432886\\
24 3.83357348838415\\
25 3.79747756028985\\
26 3.76145142514783\\
27 3.72644706864056\\
28 3.69245246189513\\
29 3.65808075464394\\
30 3.62357716874536\\
31 3.59144492928462\\
32 3.55912024117776\\
33 3.52660680582109\\
34 3.4955756168683\\
35 3.46436267211925\\
36 3.43297127659237\\
37 3.40140461725056\\
38 3.37056598182304\\
39 3.34024055862957\\
40 3.31031197581769\\
41 3.28056084504728\\
42 3.25189001698798\\
43 3.22306215419647\\
44 3.19436472297221\\
45 3.16562895443615\\
46 3.13685514578275\\
47 3.10804358963409\\
48 3.07918669941443\\
49 3.05026361234068\\
50 3.02130390458381\\
51 2.99230784741311\\
52 2.96327570810854\\
53 2.93420775004801\\
54 2.90510423279229\\
55 2.87596541216737\\
56 2.84679154034446\\
57 2.81758286591781\\
58 2.78833963398024\\
59 2.75906208619663\\
60 2.72975046087529\\
61 2.70040499303748\\
62 2.67102591448488\\
63 2.64161345386536\\
64 2.61280749301834\\
65 2.58401299806323\\
66 2.55618692910161\\
67 2.52880326608262\\
68 2.50171198354812\\
69 2.47478168107681\\
70 2.44834369544922\\
71 2.42187240555032\\
72 2.39610311008686\\
73 2.3709189898564\\
74 2.34643971393495\\
75 2.32218465203166\\
76 2.29880660586706\\
77 2.27540019465439\\
78 2.25196558141296\\
79 2.22850292729899\\
80 2.20501239163734\\
81 2.18149413195255\\
82 2.15794830399921\\
83 2.13437506179159\\
84 2.11077455763268\\
85 2.08714694214254\\
86 2.06349236428602\\
87 2.03981097139993\\
88 2.01610290921956\\
89 1.99236832190468\\
90 1.97117499663008\\
91 1.94995542991497\\
92 1.92870976134278\\
93 1.90743812901987\\
94 1.88614066959884\\
95 1.86481751830132\\
96 1.84346880894039\\
97 1.82209467394243\\
98 1.80069524436859\\
99 1.77927064993579\\
100 1.75782101903728\\
101 1.73634647876281\\
102 1.71484715491839\\
103 1.69332317204562\\
104 1.67177465344067\\
105 1.65020172117286\\
106 1.64339900008443\\
107 1.63402863443081\\
108 1.62299083681912\\
};
\addlegendentry{$I(\Sen ; \Rel^{(k)})$ generated by \cite[Algorithm~1]{PF2014} for PF};

\addplot [
color=blue,
solid,
line width = 1.5pt
]
table[row sep=crcr]{
1 7.33331134320244\\
2 7.26395610328778\\
3 7.20837774420921\\
4 7.14988072698615\\
5 7.08884392436342\\
6 7.02555895517106\\
7 6.96025726157731\\
8 6.90583692816143\\
9 6.8504558231245\\
10 6.79417284530297\\
11 6.73704008544462\\
12 6.67910394268483\\
13 6.6204060094475\\
14 6.56098378187437\\
15 6.50087123677741\\
16 6.44009930506546\\
17 6.37869626387522\\
18 6.31668806414288\\
19 6.25409860638059\\
20 6.20808101751098\\
21 6.16182107680303\\
22 6.11532462520469\\
23 6.06859722871784\\
24 6.02164419736692\\
25 5.97447060246192\\
26 5.92708129234299\\
27 5.87948090677022\\
28 5.83167389010114\\
29 5.78366450338097\\
30 5.73545683545563\\
31 5.68705481320413\\
32 5.63846221097599\\
33 5.5896826593094\\
34 5.54071965299728\\
35 5.49157655856131\\
36 5.44225662118697\\
37 5.39276297116763\\
38 5.3430986299001\\
39 5.29326651547021\\
40 5.24326944786279\\
41 5.19311015382708\\
42 5.1427912714256\\
43 5.09231535429191\\
44 5.04168487562003\\
45 4.9909022319064\\
46 4.93996974646337\\
47 4.88888967272118\\
48 4.83766419733434\\
49 4.78629544310654\\
50 4.73478547174714\\
51 4.68313628647136\\
52 4.63134983445485\\
53 4.57942800915287\\
54 4.52737265249329\\
55 4.47518555695171\\
56 4.42286846751672\\
57 4.39326034732951\\
58 4.36362036246389\\
59 4.3339487185013\\
60 4.30424561838744\\
61 4.27451126248268\\
62 4.24474584861115\\
63 4.21494957210855\\
64 4.18512262586884\\
65 4.15526520038974\\
66 4.1253774838171\\
67 4.09545966198819\\
68 4.06551191847392\\
69 4.03553443462008\\
70 4.0055273895875\\
71 3.97549096039136\\
72 3.94542532193947\\
73 3.91533064706967\\
74 3.88520710658641\\
75 3.85505486929637\\
76 3.82487410204333\\
77 3.79466496974223\\
78 3.76442763541236\\
79 3.73416226020996\\
80 3.70386900345987\\
81 3.67354802268665\\
82 3.64319947364487\\
83 3.61282351034881\\
84 3.58242028510147\\
85 3.55198994852289\\
86 3.52153264957794\\
87 3.49104853560341\\
88 3.46053775233461\\
89 3.43000044393129\\
90 3.39943675300308\\
91 3.36884682063434\\
92 3.33823078640854\\
93 3.30758878843202\\
94 3.27692096335736\\
95 3.24622744640623\\
96 3.21550837139168\\
97 3.1847638707401\\
98 3.15399407551265\\
99 3.12319911542623\\
100 3.0923791188741\\
101 3.06153421294602\\
102 3.03066452344798\\
103 2.99977017492159\\
104 2.96885129066302\\
105 2.9379079927416\\
106 2.90694040201803\\
107 2.87594863816227\\
108 2.86914591707383\\
109 2.85977555142022\\
110 2.85297283033178\\
111 2.84617010924335\\
112 2.83936738815491\\
113 2.82999702250129\\
114 2.82319430141286\\
115 2.81382393575924\\
116 2.8070212146708\\
117 2.80021849358237\\
118 2.79084812792875\\
119 2.78404540684031\\
120 2.77724268575188\\
121 2.77043996466344\\
122 2.76363724357501\\
123 2.75426687792139\\
124 2.7432290803097\\
125 2.73095139162114\\
126 2.71768562791403\\
127 2.71088290682559\\
128 2.70408018573716\\
129 2.69470982008354\\
130 2.68790709899511\\
131 2.67853673334149\\
132 2.6674989357298\\
133 2.65522124704124\\
134 2.64841852595281\\
135 2.64161580486437\\
136 2.63224543921075\\
137 2.62120764159907\\
138 2.6089299529105\\
139 2.60212723182207\\
140 2.59275686616845\\
141 2.58595414508002\\
142 2.5765837794264\\
143 2.56978105833796\\
144 2.56297833724953\\
145 2.55360797159591\\
146 2.54680525050748\\
147 2.53743488485386\\
148 2.52639708724217\\
149 2.51411939855361\\
150 2.50731667746517\\
151 2.49794631181156\\
};
\addlegendentry{$I(\Pub ; \Rel^{(k)})$ generated by \cite[Algorithm~2]{PF2014} for IB};

\end{axis}
\end{tikzpicture}% 
}
	\caption{The convergence of $I(\Sen;\Rel^{(k)})$ and $I(\Pub;\Rel^{(k)})$ for \cite[Algorithm~1]{PF2014} and \cite[Algorithm~2]{PF2014}, respectively, on the Hungarian heart disease data set in \cite{UCI2007} with $\Sen = \Set{\op{`age', } \op{`sex'}}$ and $\Pub = \Set{\op{`sex', } \op{`cholesterol'} }$.}
	\label{fig:ConvergeCompare}
\end{figure}
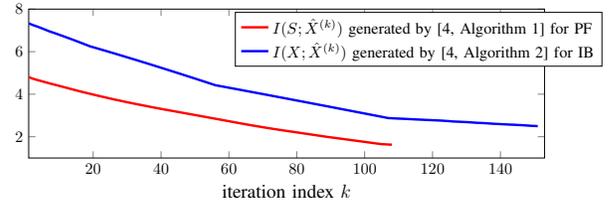

The IAC-MDSF algorithm in Algorithm~\ref{algo:CombAggloCluster} and \cite[Algorithms~1 and 2]{PF2014} all ensure a local convergence. However, \cite[Algorithms~1 and 2]{PF2014} may become very cumbersome for large $\PUB$. Fig.~\ref{fig:ConvergeCompare} shows an example of the convergence performance of \cite[Algorithms~1 and 2]{PF2014}, where both algorithms require more than $100$ iterations. In this case, \cite[Algorithm~1]{PF2014} merges $\PUB$ with $|\PUB| = 197$ into $|\REL^{(108)}| = 90$ clusters finally. Since $|\REL^{(k)}|$ is reduced by $1$ each time by a brute-force search over all $O(|\REL^{(k)}|^2)$ pairs of elements in $\REL^{(k)}$, the overall computation is around $107 \times 197^2$ large (The exact complexity is $\sum_{i =90}^{197} \frac{i(i-1)}{2}$ operations).
On the other hand, Algorithm~\ref{algo:CombAggloCluster} searches the optimal merge in the power set $2^{\REL^{(k)}}$ and allows more than $1$ reduction of $|\REL^{(k)}|$ each time so that it is able to converge only in a few iterations, e.g., Fig.~\ref{fig:ObjConverge_Health}.

The SubM-SuperM algorithm \cite{SubSuper2005} implemented in this paper for solving the MDSF problem is a greedy iterative method, where each iteration calls the min-norm algorithm \cite{MinNorm}, a submodular function minimization (SFM) algorithm that is practically fast although the asymptotic complexity is unknown.
Alternatively, one can implement \cite[Algorithm~3]{SubMDiff2012} that calls a modular function minimization algorithm with complexity $O(|\REL^{(k)}|)$ in each iteration.
In addition, MDSF is still an active research topic in combinatorial optimizations. There might be some development in this topic in the future that can be applied to Algorithm~\ref{algo:CombAggloCluster} to improve the performance (e.g., a faster convergence to a better local optimum).

\section{Conclusion}

We considered the problem of how to determine a deterministic solution $p(\rel | \pub) \in \Set{0,1}$ for the PF problem $\min_{p(\rel | \pub)} I(\Sen ; \Rel)$ s.t. $I(\Pub ; \Rel) \geq \theta_{\op{U}}$. We proposed an IAC-MDSF algorithm that generates a deterministic transition $p(\rel | \pub)$ and an alphabet $\REL$ by iteratively merging elements in $\PUB$.
Our IAC-MDSF algorithm differs from the existing algorithms in \cite{PF2014} in that it searches the optimal merge over all subsets, instead of all pairwise combinations, of the current alphabet and this problem is proved to be an MDSF, a local optimum of which could be obtained in polynomial time.
Experimental results showed that our IAC-MDSF algorithm generally outperforms the pairwise merge algorithm in \cite{PF2014} in much fewer iterations.

While the IAC-MDSF algorithm only searches a deterministic solution for the PF problem, it is worth understanding in the future how to search an optimal soft transition $p(\rel | \pub) \in [0,1]$ over the probability simplex, e.g., by the deterministic annealing method \cite{DA1998}, and whether this soft solution can improve the Pareto frontiers in Fig.~\ref{fig:PvsU_Health}.
On the other hand, as explained in Section~\ref{sec:Complexity}, it would be of interest to see if we can utilize better MDSF algorithms to improve the performance and complexity of the IAC-MDSF algorithm.

%%%%%
% Appendix:
% If needed a single appendix is created by
%
%\appendix

\appendices

\section{Proof of Theorem~\ref{theo:main}}
\label{app:theo:main}

\begin{IEEEproof}
    We have \eqref{eq:Eq} hold since
     \begin{equation}
         \begin{aligned}
         & I (\Sen;\Rel^{(k)}) - I(\Sen;\Rel^{(k)}_W) \\
         & = \sum_{\sen \in \SEN} \sum_{\rel^{(k)} \in W} p(\sen,\rel^{(k)}) \Big( \log \frac{p(\sen,\rel^{(k)})}{p(\sen) p(\rel^{(k)})} - \log \frac{p(\sen,\TW)}{p(\sen) p(\TW)} \Big) \\
         & = g(W) - f(W)
         \end{aligned}  \nonumber
     \end{equation}
    so that $I(\Sen;\Rel^{(k)}_W) = I (\Sen;\Rel^{(k)}) - g(W) + f(W)$ and
    \begin{equation}
        \begin{aligned}
            & I (\Rel^{(k)};\Rel^{(k)}_W) \\
%            & = \sum_{\rel^{(k)} \in \REL^{(k)}} \sum_{\rel^{(k)}_W \in \REL^{(k)}_W}  p(\rel^{(k)},\rel^{(k)}_W) \log \frac{ p(\rel^{(k)},\rel^{(k)}_W) }{ p(\rel^{(k)}) p(\rel^{(k)}_W) } \\
                            & \  = -\sum_{\rel^{(k)} \notin W} p(\rel^{(k)}) \log p(\rel^{(k)}) - \sum_{\rel^{(k)} \in W} p(\rel^{(k)}) \log p(\TW) \\
                            & \  = H(\Rel^{(k)}) + f(W). \\
        \end{aligned}  \nonumber
    \end{equation}
    For function $l = u(t(W))$, if $t$ is a modular (both submodular and supermodular) set function such that $t(W) = \sum_{i \in W} t_i, \forall W \subseteq V$ for the vector $\mathbf{t} \in \RealP^{|V|}$ and $u \colon \Real \mapsto \Real$ is convex, $l$ is supermodular \cite[Proposition~37]{Bach2010SFMtut}. Then, $-l$ is submodular.
    Rewrite $f(W) = \sum_{\rel^{(k)} \in W} p(\rel^{(k)})  \log p(\rel^{(k)}) - p(\TW) \log {p(\TW)}$. Here, $\sum_{\rel^{(k)} \in W} p(\rel^{(k)})  \log p(\rel^{(k)})$ is modular. Since $p(\TW)$ is nonnegative and modular and $-y \log y$ is convex in $y$, $- p(\TW) \log {p(\TW)}$ is submodular. Therefore, $f$ is submodular.
    Also, for all $W \subseteq Y$, %$f(W) - f(Y) = \sum_{\rel^{(k)} \in W} p(\rel^{(k)}) \log \frac{p(\TY)}{p(\TW)} - \sum_{\rel^{(k)} \in Y \setminus W} p(\rel^{(k)}) \log \frac{p(\rel^{(k)})}{p(\TY)} \geq 0.$
    \begin{multline}
            f(W) - f(Y) = \sum_{\rel^{(k)} \in W} p(\rel^{(k)}) \log \frac{p(\TY)}{p(\TW)}  \\
            - \sum_{\rel^{(k)} \in Y \setminus W} p(\rel^{(k)}) \log \frac{p(\rel^{(k)})}{p(\TY)} \geq 0. \nonumber
    \end{multline}
    In the same way, we can prove that $g$ is submodular and nonincreasing.
\end{IEEEproof}

%
% If several appendices are needed, then the command
%
%
%
% in combination with further \section-commands can be used.
%%%%%

%\section*{Acknowledgment}
%
%We are indebted to Michael Shell for maintaining and improving
%\texttt{IEEEtran.cls}.

\bibliographystyle{IEEEtran}
\bibliography{PrivacyBIB}

%%%%%%
%% To balance the columns at the last page of the paper use this
%% command:
%%
%\enlargethispage{-1.2cm}
%%
%% If the balancing should occur in the middle of the references, use
%% the following trigger:
%%
\IEEEtriggeratref{3}
%%
%% which triggers a \newpage (i.e., new column) just before the given
%% reference number. Note that you need to adapt this if you modify
%% the paper.  The "triggered" command can be changed if desired:
%%
%\IEEEtriggercmd{\enlargethispage{-20cm}}
%%
%%%%%%

%%%%%%
%% References:
%% We recommend the usage of BibTeX:
%%
%\bibliographystyle{IEEEtran}
%\bibliography{definitions,bibliofile}
%%
%% where we here have assume the existence of the files
%% definitions.bib and bibliofile.bib.
%% BibTeX documentation can be obtained at:
%% http://www.ctan.org/tex-archive/biblio/bibtex/contrib/doc/
%%%%%%

%% Or you use manual references (pay attention to consistency and the
%% formatting style!):

\end{document}